\documentclass[lettersize,magazine]{IEEEtran}
\usepackage{amsmath,amsfonts}
\usepackage{algorithmic}
\usepackage{algorithm}
\usepackage{array}
\usepackage[caption=false,font=footnotesize,labelfont=rm,textfont=rm]{subfig}
\usepackage{multirow}
\usepackage{textcomp}
\usepackage{stfloats}
\usepackage{xcolor}
\usepackage{url}
\usepackage{verbatim}
\usepackage{graphicx}
\usepackage{cite}
\usepackage{ragged2e}
\begin{document}
\bstctlcite{setting}
\title{Wireless Network Digital Twin for 6G: Generative AI as A Key Enabler}

\author{Zhenyu~Tao,
        Wei~Xu,~\IEEEmembership{Senior Member,~IEEE},
        Yongming~Huang,~\IEEEmembership{Senior Member,~IEEE},
        Xiaoyun~Wang,
        and~Xiaohu~You,~\IEEEmembership{Fellow,~IEEE}

\thanks{Zhenyu Tao, Wei Xu, Yongming Huang, and Xiaohu You are with the National Mobile Communications Research Lab, Southeast University, Nanjing 210096, China, and also with the Pervasive Communication Research Center, Purple Mountain Laboratories, Nanjing 211111, China (email: \{zhenyu\_tao, wxu, huangym, xhyu\}@seu.edu.cn).}
\thanks{Xiaoyun Wang is with the China Mobile, Beijing 100032, China (e-mail:
wangxiaoyun@chinamobile.com).}
\thanks{Xiaohu You is the corresponding author of this paper.}
}

\markboth{IEEE wireless communications, December~2023}%
{Shell \MakeLowercase{\textit{et al.}}: A Sample Article Using IEEEtran.cls for IEEE Journals}


\maketitle

\begin{abstract}
Digital twin, which enables emulation, evaluation, and optimization of physical entities through synchronized digital replicas, has gained increasing attention as a promising technology for intricate wireless networks. For 6G, numerous innovative wireless technologies and network architectures have posed new challenges in establishing wireless network digital twins. To tackle these challenges, artificial intelligence (AI), particularly the flourishing generative AI, emerges as a potential solution. In this article, we discuss emerging prerequisites for wireless network digital twins considering the complicated network architecture, tremendous network scale, extensive coverage, and diversified application scenarios in the 6G era. We further explore the applications of generative AI, such as Transformer and diffusion model, to empower the 6G digital twin from multiple perspectives including \textcolor{black}{physical-digital modeling}, synchronization, and slicing capability. Subsequently, we propose a hierarchical generative AI-enabled wireless network digital twin at both the message-level and policy-level, and provide a typical use case with numerical results to validate the effectiveness and efficiency. Finally, open research issues for wireless network digital twins in the 6G era are discussed.
\end{abstract}


\section{Introduction}

From the first-generation (1G) of analog communication systems to the contemporary fifth-generation (5G) era, we have witnessed a rapid evolution of wireless communication networks. As we stand on the brink of the sixth-generation (6G) era, there is a continuous emergence of novel technologies aimed at meeting superior performance demands for next-generation networks, such as microsecond-scale latency, terabits-per-second-scale peak data rate, and ubiquitous connectivity \cite{you2023toward}. The integration of advanced wireless technologies, for instance, space-air-ground integrated networks, integrated sensing and communication (ISAC), and native network intelligence, significantly escalates the intricacy of network structure and functionality, necessitating digitalization of wireless networks for the sake of network performance evaluation and dynamic optimization. To tackle this issue, wireless network digital twin, which is capable of establishing a faithful replica of physical networks, has been considered as a promising technology in the 6G era for their extensive applications throughout research and development life cycles. It provides a risk-free environment for preliminary research on innovative technologies, facilitates the verification of extended 6G network architectures before deployment, and accelerates adaptation along with real-time wireless network management during practical applications.

The concept of digital twin, encompassing a physical entity, a digital representation, and a communication channel, was first proposed by Grieves in 2003 \cite{grieves2017digital}. In recent years, endeavors have been made to establish digital twins of wireless networks through diversified paradigms, including programming, mathematical modeling, and artificial intelligence (AI). These developments empower network operators to predict potential issues and seek optimization solutions via digital twin-based emulation and evaluation. In \cite{rodrigo2023digital}, digital twins have been employed for assessing performance of a core network, forecasting service changes, and optimizing network management. In the realm of radio access network (RAN), the digital twin has been established to facilitate intelligent resource management for the RAN \cite{zhang2023digital}. For wireless network topologies, graph neural network-based digital twins have been developed to enhance the latency prediction across potential topologies \cite{wang2020graph}. However, creating digital twins is usually nontrivial, necessitating not only accurate modeling of network functions but also efficient synchronization between the physical and digital entities. This poses challenges for employing current digital twin methods in the evolving 6G network. Additionally, existing wireless network digital twins lack adaptability and scalability to cope with dynamic network status and user demands.

The impending vision of 6G era, as released by the international telecommunication union (ITU) \cite{ITU2023} in June 2023, will inevitably result in a wireless network with substantial scale, intricate structures, extensive states, and diversified services. To achieve network autonomy and intelligence, it is widely advocated that AI should be deeply integrated into wireless networks, encompassing the paradigms of AI for network (AI4Net) and network for AI (Net4AI). Currently, efforts have been made towards developing AI-enhanced network digital twins, utilizing data-driven neural networks to simulate the network function, network topology, and user behavior within wireless networks \cite{tao2023}. In addition, deep reinforcement learning (DRL) has been an effective approach to integrate with wireless network digital twins to tackle optimization tasks like network admission control and resource allocation \cite{zhang2023digital}. However, current digital twin methods still face challenges in the 6G era in terms of \textcolor{black}{physical-digital modeling}, synchronization, and slicing capability. Consequently, there is an immediate need to explore innovative approaches in wireless network digital twins to align with the evolving requirements of 6G. 

The rapid development of deep learning (DL) technologies has facilitated the emergence of generative AI models, capable of creating novel and realistic content such as text and images. With available large-scale data, powerful computing resources, and novel algorithms, generative AI has fortunately been achieving successful commercialization in various domains. ChatGPT, as a conversational agent based on a generative Transformer decoder, generates fluent and engaging responses according to input sentences from users. In terms of images, DALL·E based on generative diffusion models, has shown the ability to generate high-quality and diverse images from scratch and conditioned on a given input. The achievements in domains of natural language processing (NLP) and computer vision (CV) have galvanized researchers to integrate generative AI into mobile networks \cite{xu2023unleashing}. Similarly, these successes have inspired us to leverage generative AI in the enhancement and substitution of digital twin techniques in the 6G era. The generative characteristics of these AI models hold promise in addressing the issue of data scarcity encountered during the construction of wireless network digital twins, while their superior transferability could be instrumental in enabling digital twins to rapidly adapt to evolving scenarios in 6G. Furthermore, the advanced and diversified architectures of generative AI models offer potentials in creating digital twins for wireless networks from multiple dimensions, thereby constructing a more comprehensive hierarchical wireless network digital twin.

\begin{figure*}[!t]
\centering
\includegraphics[width=1\textwidth]{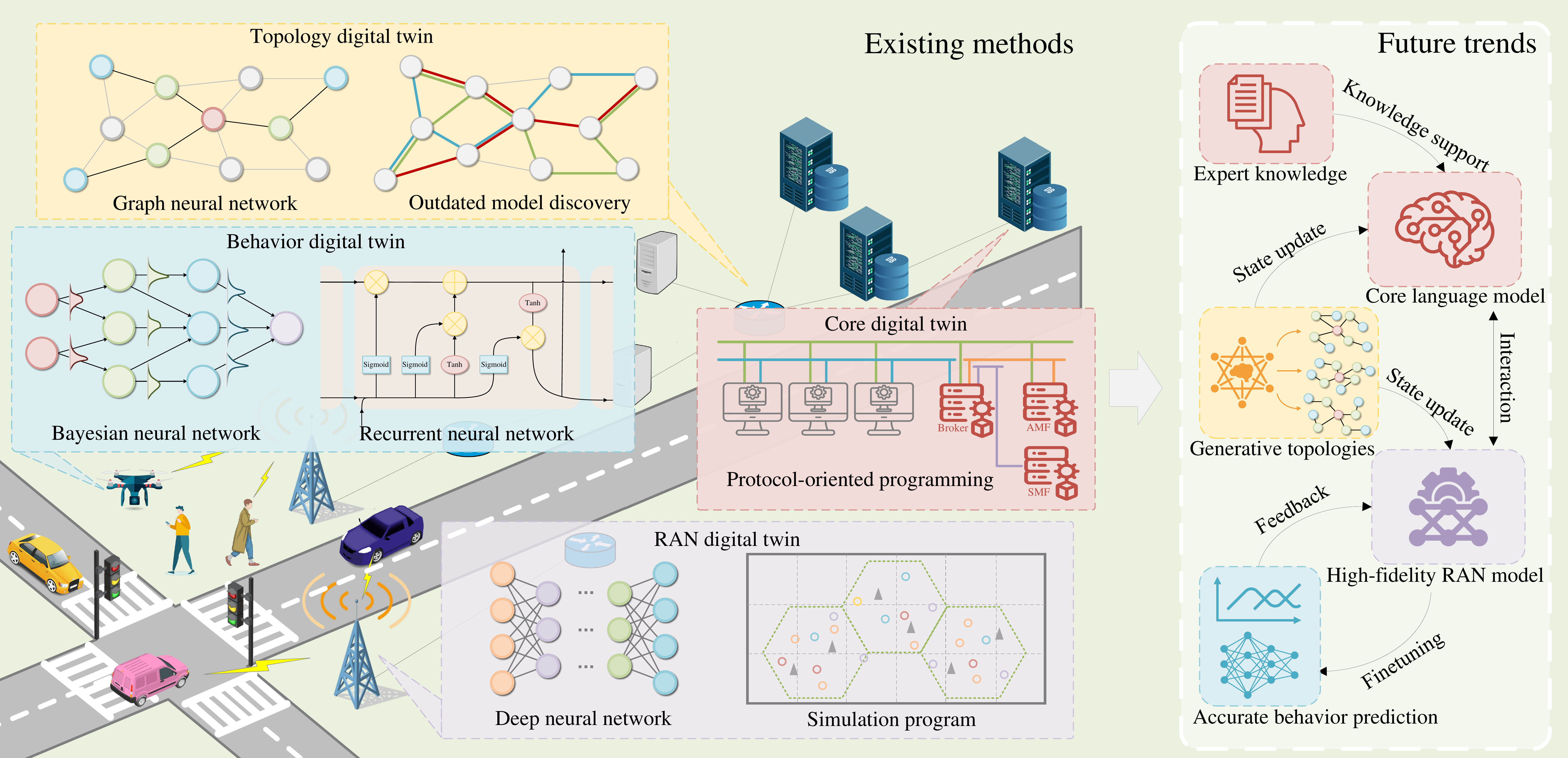}
\caption{Digital twin solutions for wireless network.}
\label{fig1}
\end{figure*}

The remainder of this article is organized as follows. First, we introduce wireless network digital twins for 6G networks. Then, we analyze potential applications of generative AI in the 6G network digital twins. A hierarchical generative AI-enabled wireless network digital twin is presented, followed by a case study with numerical verifications. Finally, future research directions are discussed before a conclusion.

\section{Wireless Network Digital Twin in 6G Era}
\subsection{Tractable Solutions of Wireless Network Digital Twins}
\textcolor{black}{Digital twin has been recognized as a vital facilitator in wireless networks \cite{you20236G}. Due to the intricacy of modern wireless networks, however, it is usually challenging to establish a single digital twin for an entire wireless network. Consequently, a common tractable solution is to segment the wireless network and create digital representations for individual components such as the core network, network topology, RAN, and user behaviors, as depicted in Fig. \ref{fig1}.}

\textcolor{black}{In practice, digital twins can be developed selectively for these network components to meet individual objectives. For instance, a dedicated RAN digital twin can be created for providing a pre-verification environment for intelligent resource management \cite{zhang2023digital}. Similarly, accurate traffic prediction necessitates a digital twin that emulates behaviors of network entities including mobile terminals and infrastructure nodes \cite{ruah2023bayesian}. }

\textcolor{black}{Modeling methodologies for these digital twins are customized based on the architecture and functionality of their respective network elements. Currently, digital twins for core networks are primarily developed through conventional programming according to protocols, with the integration of open-source projects to streamline the development process \cite{rodrigo2023digital}, while network topology digitalization usually exploits graph neural networks due to their structural similarity \cite{wang2020graph}. Principal component analysis (PCA)-based outdated model discovery has also been employed in construction of network topology digital twin \cite{jia2023new}. A summary of existing digital twin solutions for wireless networks is provided in Table \ref{tab1}.}

\renewcommand\arraystretch{1.5}
\begin{table*}[]
\centering
\caption{\textcolor{black}{Digital Twins of Different Wireless Network Elements}}
\label{tab1}
\begin{tabular}{|>{\Centering\arraybackslash}p{2 cm}|>{\Centering\arraybackslash}p{2.8cm}|>{\Centering\arraybackslash}p{3.8cm}|>{\Centering\arraybackslash}p{4 cm}|>{\Centering\arraybackslash}p{3 cm}|}
\hline
Network element & Modeling strategy & Enabling technology & Objectives&Reference\\
\hline
\multirow{2}{*}{Core network} & \multirow{2}{*}{Model-driven} & \multirow{2}{*}{Protocol-oriented programming} & Testbed, risk control, real-time optimization & \multirow{2}{*}{M. S. Rodrigo \emph{et al.} \cite{rodrigo2023digital}} \\
\cline{1-5}
\multirow{3}{*}{RAN} & \multirow{2}{*}{Model-driven} & \multirow{2}{*}{Simulation program} & Optimal capacity-sharing for network slicing & \multirow{2}{*}{I. Vilà \emph{et al.} 2023} \\
\cline{2-5}
 & \multirow{2}{*}{Data-driven} & \multirow{2}{*}{Deep neural network} & Resource management in networks slicing & \multirow{2}{*}{Z. Zhang \emph{et al.} \cite{zhang2023digital}}  \\
\hline
\multirow{2}{*}{Topology} & Model-driven &  Outdated model discovery &   Spatial-temporal load balance & P. Jia \emph{et al.} \cite{jia2023new} \\
\cline{2-5}
 & Data-driven & Graph neural network &  End-to-end latency prediction &  H. Wang \emph{et al.} \cite{wang2020graph}\\
\cline{1-1} \cline{2-5}
\multirow{3.5}{*}{Behavior} & \multirow{3.5}{*}{Data-driven} & \multirow{2}{*}{Bayesian neural network} &  Traffic prediction, anomaly detection, data collection & \multirow{2}{*}{C. Ruah \emph{et al.} \cite{ruah2023bayesian}}\\
\cline{3-5}
 &  & \multirow{2}{*}{Recurrent neural network} &  Cache monitoring, packet positioning & \multirow{2}{*}{G. Lin \emph{et al.} 2022} \\
\hline 
\end{tabular}
\end{table*}


\subsection{Requirements of Wireless Network Digital Twins for 6G}
While these methods have successfully established digital twins for current wireless networks, the advent of 6G era comes with multifarious stringent requirements that are far beyond the capabilities of existing methods. Besides typical 5G network key performance indicators (KPIs) in terms of latency, data rate, mobility, and so on, additional indicators such as global coverage, applicable AI-related capabilities, and sustainability have been raised by ITU in the international mobile telecommunications 2030 (IMT-2030) framework \cite{ITU2023} to accommodate disruptive use cases and applications in 6G. In order to satisfy the emerging KPIs, numerous innovative technologies and network architectures have been proposed and incorporated into wireless networks, including representatives as integrated sensing, computing, and communication \cite{xu2023edge}, autonomous networks, and native AI \cite{we2022ns}. This surge of innovations has substantially increased \textcolor{black}{network complexity, scalability, and heterogeneity,} necessitating advancements in corresponding digital twin technologies. These requirements are broadly categorized into three key areas.

\textbf{\textcolor{black}{Physical-digital modeling}:}
While model-driven digital twin methods can effectively emulate network functions in many cases, the substantial time and capital investment for system replication renders them unsuitable in this fast-paced technological era. Furthermore, the heterogeneity of wireless network architectures and functions in varying 6G scenarios also makes it impractical to conduct extended programming and customization for each type of network during the establishment of digital twins. Hence, it is an inevitable path of self-monitored and automated approaches, for instance, AI-enabled methods, to take over the conventional model-driven method to cater to the demands in the 6G era. 

On the other hand, data-driven methodologies, which employ neural networks to intelligently establish digital twins, are inherently data-hungry. These AI approaches require large volumes of data to ensure learning convergence, especially for the emerging massive wireless networks. With the expected surge in connection density, network scale, and coverage in 6G, it is predicted to face a substantially and potentially exponentially growing need for training data. However, data collection, filtering, and labeling in physical networks are in fact costly, time-consuming, and privacy-sensitive. It is definitely impossible to acquire data covering all potential scenarios and conditions that the network may encounter. Consequently, a tractable and efficient strategy for data acquisition and dataset construction becomes a fundamental challenge to facilitate the training of digital twin models via data-driven methods.

\textbf{Physical-digital synchronization:}
In order to achieve accurate replication of a physical 6G network, timely synchronization between the physical and digital entities is a key ingredient to maintain synchronized operations of digital twins. In current 5G networks, synchronization is achieved through a local deployment of digital twins. \textcolor{black}{However, the advent of the 6G era significantly amplifies the scale and complexity of wireless networks, thereby posing challenging requirements for wireless network digital twins. For example, the space-air-ground integrated 6G networks extend two-dimensional (2D) terrestrial coverage to three-dimensional (3D) global coverage, incorporating terrestrial and aerial base stations. Moreover, heterogeneous networks employ a combination of macro, pico, and femto base stations for flexible and cost-effective coverage, while edge computing necessitates collaborative optimization across various nodes including edge and cloud. For these scenarios, traditional local deployment of digital twins becomes inadequate. In this context, the construction of digital twins can either adopt a distributed deployment strategy with remote collaboration or opt for a centralized deployment within the centralized cloud, coordinating distributed segments via synchronous signaling.} Both methods consume excessive bandwidth for the overhead and could potentially interfere with regular network communications, even jeopardizing the guarantee of ultra-reliable low-latency communication in future networks. Moreover, noise and interference during wireless data transmission can result in the corruption and distortion of exchanged data, compromising the accuracy and reliability of digital twins. Therefore, it is imperative to develop an affordable transmission strategy to fulfill the stringent synchronization requirements of 6G wireless network digital twins. Given the prior success of semantic communication \cite{xu2023edge} and AI-driven physical-layer communication \cite{qin2019deep}, AI technology is supposed to be a viable solution for meeting these requirements.

\textbf{Slicing capability:}
Network slicing for 6G enables the creation of multiple distinct virtual networks on top of a shared physical hardware infrastructure, allowing dynamic services and applications to have customized network structures, functions, procedures, and resources with performance guarantees. There are three typical types of slices in the current 5G network: enhanced mobile broadband (eMBB), ultra-reliable low-latency communication (URLLC), and massive machine-type communication (mMTC). However, existing digital twin methods exhibit limitations in providing customized functionality for each slice type and establishing digital slices in alignment with the sliced physical network. As we move into the 6G era, the number and diversity of slices are expected to further increase to accommodate new use cases and scenarios. The typical scenarios, previously as eMBB, URLLC, and mMTC in 5G, have evolved to immersive communication, hyper-reliable low-latency communication (HRLLC), and massive communication, respectively, along with the addition of ubiquitous connectivity, AI and communication, and ISAC \cite{ITU2023}. Although various network slicing technologies have successfully integrated AI techniques like deep learning and reinforcement learning for rapid adaptation to emerging scenarios in 6G \cite{we2022ns}, such paradigms have not yet been extensively adopted in the current network digital twin solutions. Consequently, introducing the slicing capability to wireless network digital twins becomes increasingly significant for addressing the dynamic visions and application scenarios in 6G. 

\textcolor{black}{Among all these requirements, physical-digital modeling is considered the primary challenge in the 6G era. It serves as the foundation for wireless network digital twins, directly impacting their fidelity in representing physical network elements and their effectiveness in network emulation, evaluation, and optimization. Moreover, the evolving scalability and heterogeneity of the impending 6G network present additional obstacles in the development of modeling techniques.}

\section{Potentials of Generative AI in 6G Wireless Network Digital Twin}
Generative AI models, typically Transformer, generative adversarial network (GAN), and diffusion model, provide opportunities to fulfill the evolving requirements of 6G wireless network digital twins. We explore applications of these generative AI technologies to enable 6G digital twins from four perspectives.

\begin{figure*}[t]
\centering
\includegraphics[width=1\textwidth]{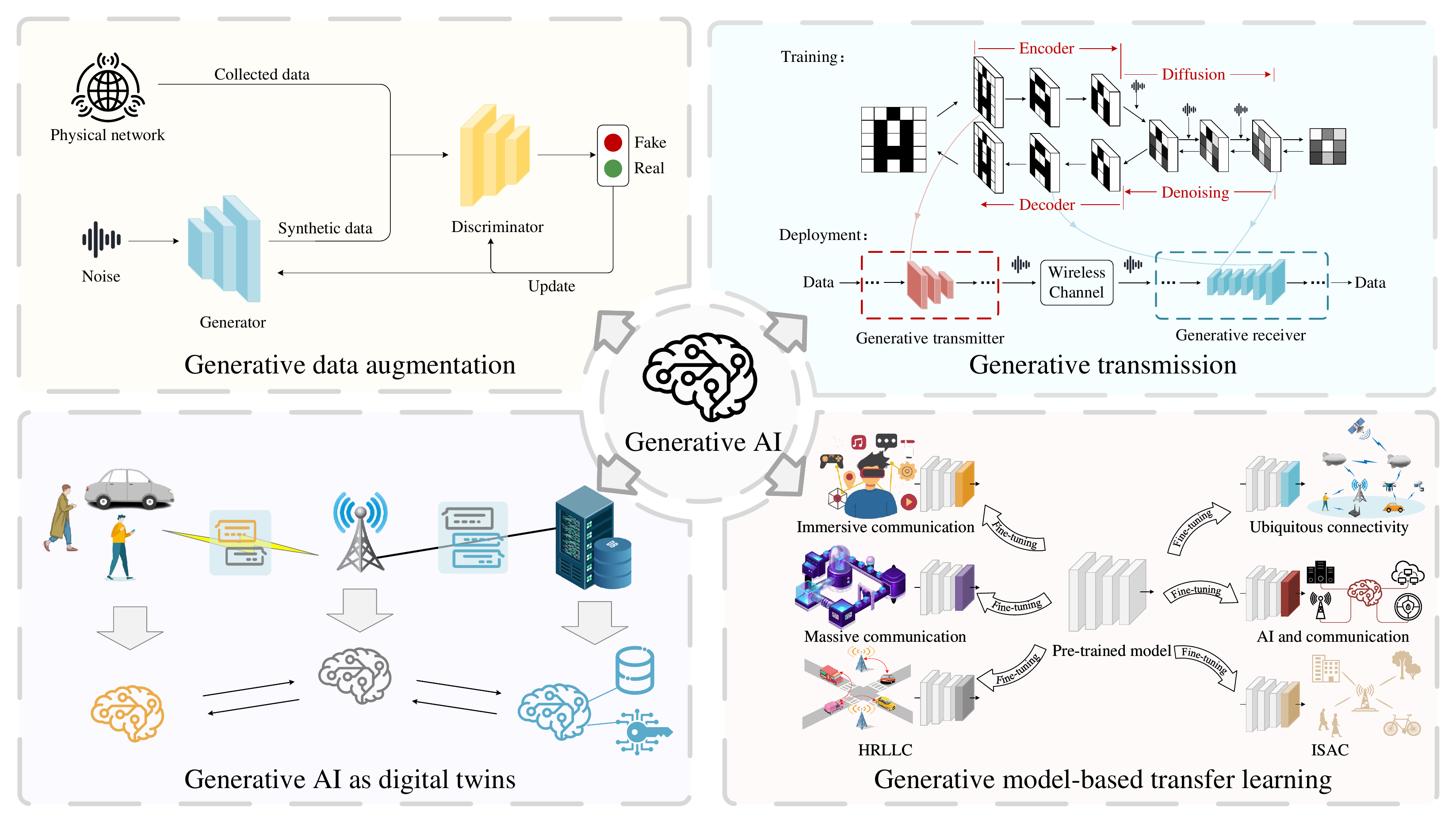}
\caption{\textcolor{black}{Generative AI in 6G wireless network digital twin.}}
\label{fig2}
\end{figure*}

\subsection{Generative Data Augmentation for Digital Twin}
Compared to existing 5G digital twins, the data-driven \textcolor{black}{modeling} method for wireless network digital twins is predicted to be much more data-hungry in the 6G era. Data augmentation, a technique that enhances both data quality and quantity by generating new samples from existing ones, is a common solution. GAN is a classic generative AI model that consists of two competing neural networks: a generator producing fake data samples, and a discriminator distinguishing between real and fake data samples. Through the alternating training of both models, GAN gradually acquires the capability to generate synthetic data samples indistinguishable from real ones, making it a valuable tool for data augmentation, particularly in the intricate 6G network. In the 6G era, network services have been further categorized into more segments, posing difficulties for the acquisition of specific network service data. The application of generative data augmentation offers a feasible solution by creating diverse synthetic network data, such as user behaviors and traffic patterns, based on collected real-world data. In addition, through generative data augmentation, some possible but previously unencountered network topology and network state can be generated. Generative data augmentation enhances not only the volume but also the diversity of wireless network data, thereby facilitating the training of wireless network digital twin models and enabling the assessment of potential risks in 6G networks.

\subsection{Generative Transmission for Digital Twin}

\textcolor{black}{Wireless network digital twins can leverage innovative generative AI techniques, such as the diffusion model, to enhance the transmission process. The diffusion model involves two main phases: a forward diffusion process that progressively introduces noise to data until it reaches a predefined noise level, and a reverse denoising process that gradually eliminates the noise until the data returns to its origin. This empowers the diffusion model to effectively learn from highly noisy data and generate high-fidelity samples. By combining classic encoder-decoder models, such as the U-net and the deep joint source-channel coding model, it can learn the process of source compression, noise introduction, signal denoising, and data reconstruction in wireless communication in an offline manner.}

\textcolor{black}{After thorough training, the encoder serves as a generative transmitter for the synchronization of 6G wireless network digital twins. It enables remote collaboration among distributedly deployed digital twins, and facilitates physical-digital synchronization between networks and digital twins deployed in a central cloud. The generative transmitter extracts key information from the synchronization data and compresses them into a low-dimensional latent space. After passing through a noisy channel, the generative receiver, consisting of the reverse diffusion model and decoder, restores the noisy data to its original state. Additionally, the transmission performance can be further enhanced for online training by utilizing real-time data collected from 6G wireless network digital twins. Such a generative transmission solution can substantially facilitate the physical-digital synchronization of 6G wireless network digital twins, especially in complex scenarios like the space-air-ground integrated communication in the 6G era.}

\subsection{Generative AI as Digital Twins}
Apart from assisting digital twin construction, a generative AI model itself can serve as a digital twin. These generative models replicate the behavior of physical networks, offering a more efficient alternative to labor-intensive and time-consuming model-driven solutions. By viewing the interaction signaling messages between different network entities as text, generative language models, for example, Transformers, can be employed in establishing a message-level digital twin. Transformer utilizes self-attention mechanisms to capture extended dependencies and contextual information from sequential data. Through training with extensive message data, the Transformer-based digital twin is able to emulate network functions by generating responses according to the requests and past interaction messages. It is worth noting that Transformers can encounter difficulties in learning specific algorithms like encryption, due to the lack of prior knowledge, including network user identity document (ID), flow billing information, and so on. Thus, a faithful message-level digital twin for the 6G network can be established through cooperations among Transformer-based models, databases, and dedicated algorithms.

\subsection{Generative Model-based Transfer Learning for Digital Twin}
To tackle diverse scenarios and heterogeneous network structures in 6G, the slicing capability also needs to be involved in digital twins. For instance, compared to the massive communication scenario, the wireless network digital twins for HRLLC need to ensure higher quality of service, conduct more concise service request procedures, and serve relatively fewer end users. These differences are reflected in the composition and distribution of the constructed dataset, such as behaviors, network topologies, and exchanged messages. To deal with this issue, customized digital twins for diverse slices can be achieved through generative model-based transfer learning. By exploiting knowledge from a well-trained source task, transfer learning helps the 6G wireless network digital twin to avoid training generative AI models from scratch and to accelerate the adaptation to new, but similar, scenarios. Particularly, after training in a general scenario, most generative AI-based digital twins implement transfer learning through parameter sharing and retraining on newly acquired data, thereby preserving their effectiveness in new scenarios. However, such a method might not suffice for large generative models like Transformers and diffusion models. In such a case, fine-tuning is ready to be employed by freezing parameters in lower layers of the pre-trained model of a digital twin, while adjusting parameters in the last few layers based on new data in changing 6G applications. It allows the digital twin to rapidly fine-tune its representations and predictions in different slices while preserving the general functionality.


\section{Hierarchical Generative AI-Enabled Wireless Network Digital Twin: A Case Study}

\begin{figure*}[!t]
\centering
\includegraphics[width=1\textwidth]{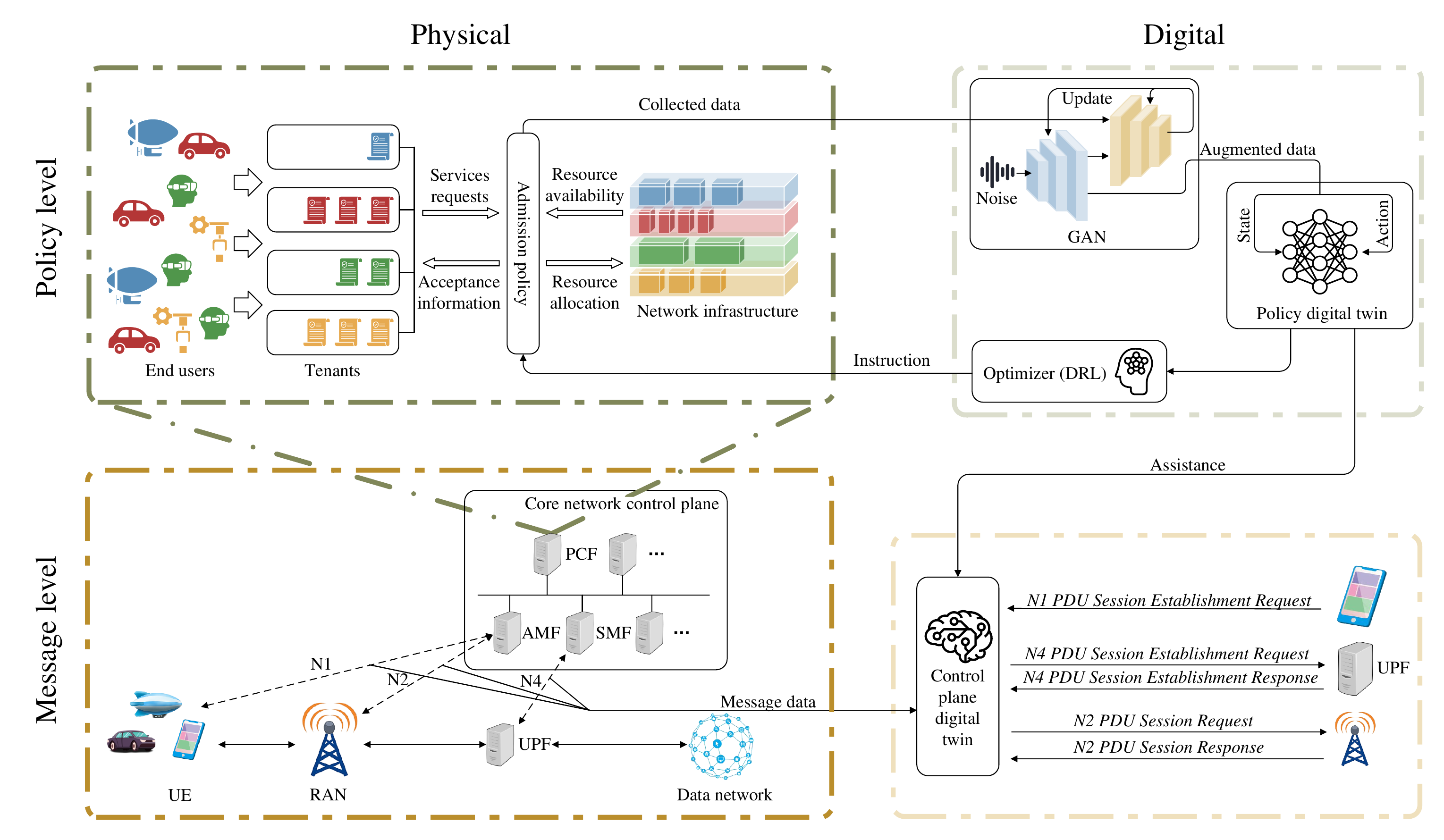}
\caption{A framework of hierarchical generative AI-enabled wireless network digital twin.}
\label{fig3}
\end{figure*}

Given the current absence of a definitive 6G network architecture, we consider a typical 5G wireless network consisting of a data network, a core network, RAN, and user equipments (UEs). With the technological evolution of network function virtualization (NFV) and software-defined network (SDN), the 5G network has introduced control and user plane separation (CUPS). In this architecture, the user plane concentrates solely on managing the substantial user traffic from UE to the data network. In contrast, the control plane, which integrates most of the core network functions, uses a relatively small number of critical signaling messages to govern the entire user plane. To realize the digitalization of core network and support the assessment, development, and optimization of network technologies, we establish a hierarchical generative AI-enabled digital twin of a beyond 5G (B5G) core network control plane, as depicted in Fig. \ref{fig3}. 

\subsection{Hierarchical Framework}
The hierarchical digital twin consists of two levels, that is, a message-level digital twin based on external exchanged messages, and a policy-level digital twin utilizing internal network states and actions.

The message-level digital twin is established upon a generative Transformer model, trained through signaling messages between the control and user planes. The core network is composed of multiple virtualized network functions, including access and mobility management function (AMF), session management function (SMF), policy control function (PCF), user plane function (UPF), and many others. Control signaling messages between the control and user planes flow through distinct interfaces, which include the N1 interface between UE and AMF, the N2 interface between RAN and AMF, and the N4 interface between SMF and UPF. Therefore, by capturing interaction messages from these interfaces and converting them into sequential directed dialogue texts, we construct a message-level dataset. This dataset is then utilized in training a message-level digital twin to emulate the functions of the core network control plane \cite{tao2023}.

While the purely message-based digital twin successfully replicates network functions of the control plane in conventional situations, it may experience malfunctions in certain scenarios due to the lack of prior knowledge. For instance, the successful establishment of a session for a specific UE depends on not only the UE's legitimacy but also current network conditions such as queueing requests from other UEs and remaining resources. The generative Transformer model can hardly process the entire history of interaction signaling messages simultaneously or extract network state information from these messages. As a solution, the policy-level digital twin is introduced to replicate internal network policies \cite{tao2023digital}, such as admission control, resource allocation, and session management, and assist the operation of the message-level digital twin. 

In order to establish the policy-level digital twin, the network state and action should be collected whenever a decision is made under a specific network policy, thereby constructing the training dataset. Considering the scarcity and rarity of data, a generative AI-based data augmentation method is employed to enrich the training dataset. We exploit a GAN model in which the generator takes a random sample from the entire state space as input and generates an action within the action space as output, while the discriminator distinguishes between the state-action pair from the actual network policy and the synthetic state-action pair. After training, this GAN model can construct and generate a sufficient augmented dataset, which is used to train the policy-level digital twin model composed of multilayer neural networks through supervised learning. 

The policy-level digital twin is capable of replicating the actual network policy in the physical network, thereby assisting the message-level digital twin in accurately emulating core network functions, particularly in scenarios involving policy-related messages. Furthermore, in response to the growing demand for 6G network optimization, the policy-level digital twin can be leveraged within DRL to enhance the existing policy of the physical network. \textcolor{black}{Specifically, the policy-level digital twin, which has parameterized the network policy through supervised learning, serve as the actor network in the actor-critic style DRL model, for example, the advantage actor critic (A2C). By tailoring rewards according to specific objectives, such as revenue maximization or fairness optimization, we can iteratively optimize the network policy to meet desired goals.}

\subsection{The Case Study}
In the case study, we consider a specific admission control policy in the policy-level digital twin. In the admission control process, a sliced network with four distinct 6G slices is considered, including slices for immersive communication, HRLLC, massive communication, and ubiquitous connectivity. As illustrated in the upper left of Fig. \ref{fig3}, various service requests arise from end users, and these requests are directed to tenants responsible for requisitioning and allocating network infrastructure resources to fulfill their service needs. The admission policy analyzes the current queue of requests and the available resources to determine the viability and priority of accepting a specific request. Service requests that have not been admitted remain in the queue, awaiting admission until they expire due to a timeout.

\begin{figure}
    \centering
    \subfloat[Message prediction performance]{\includegraphics[width=0.48\textwidth]{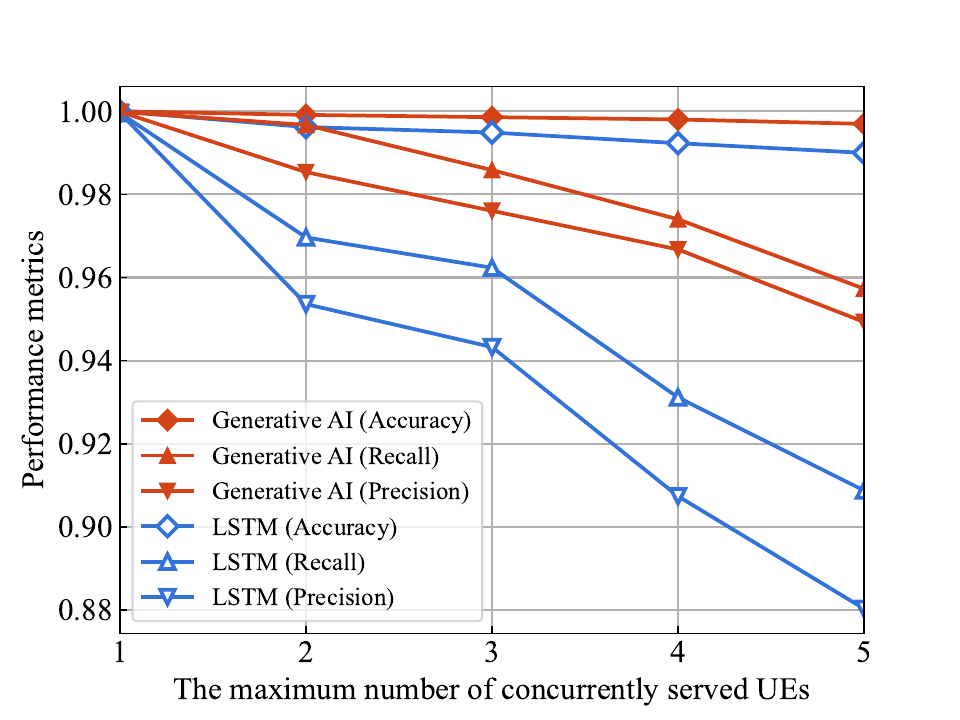}
    \label{fig4}
    }\\
    \subfloat[Policy optimization performance]{\includegraphics[width=0.48\textwidth]{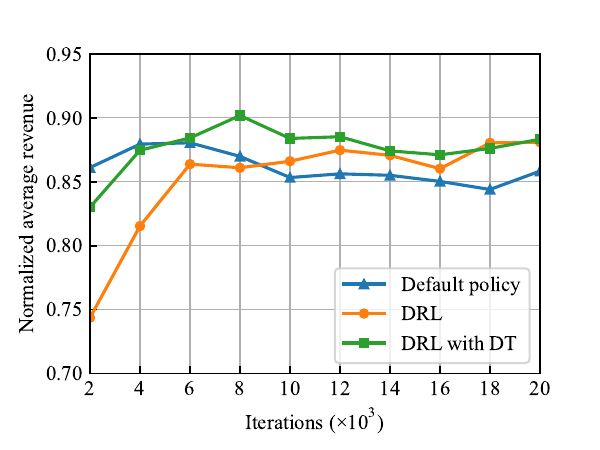}
    \label{fig5}}
    \caption{\textcolor{black}{Performance comparison for wireless network digital twins}}
\end{figure}

We set three types of resources for the network services: radio, computing, and storage resources, with resource utilization varying across services within the network slices. For instance, the immersive communication slice predominantly utilizes radio resource, whereas the HRLLC slice places a heavier demand on computing resource. Arrival rates for different service requests are customized according to the slice feature and following a Poisson arrival process. Unique means for service times in different slices are also set, following an exponential distribution. The admission policy in the physical network is implemented as a greedy algorithm. In terms of models, the performance of the generative Transformer model and a conventional non-generative long short-term memory (LSTM) model are compared. \textcolor{black}{The maximum number of concurrent UEs is altered to assess the robustness of the two models across various scenarios. Accuracy serves as a fundamental performance metric. Considering the sparsity of the information in signaling messages, we also adopt the recall and precision metrics to better evaluate the prediction of positive samples, which refer to the information elements carried in the messages.} \textcolor{black}{In Fig. \ref{fig4}, the generative AI-enhanced digital twin model exhibits notable performance superiority against the non-generative LSTM model, especially in scenarios with a high volume of concurrently served UEs. This superiority is creditable to the parallel architecture and the distinctive attention mechanism in the Transformer. The parallel architecture enables the model to process all historical signaling messages simultaneously, avoiding the long-term dependency issue, while the attention mechanism helps it concentrate on the most important messages, particularly those belonging to the currently processed UE. Additionally, to evaluate the optimization efficacy of wireless network digital twins, we conducted an experiment employing the optimization strategy outlined in the previous section. The results, as depicted in Fig. \ref{fig5}, illustrate a significant advantage of the digital twin-based DRL compared to the conventional DRL, particularly within the first thousands of iterations.}

\section{Open Opportunities and Issues}
In the following, we discuss open research opportunities and issues pertaining to generative AI-enabled 6G wireless network digital twins.

\subsection{\textcolor{black}{Cross-layer Digital Twin Collaboration}}
\textcolor{black}{While the majority of digital twin methods are tailored for specific network elements such as RAN and core networks, end-to-end optimization of wireless networks calls for the collaboration of digital twins across multiple network elements. It is therefore imperative to deeply explore the synthesis of 6G wireless network digital twins with diverse structures and functions, potentially operating asynchronously, as well as their utilization for specific optimization objectives. Besides, the substantial scale of generative AI models like Transformers opens new research opportunities in terms of improving response time and computational resources utilization, while the needs for coordinating optimization goals, functions, and synchronization intervals arise with challenging issues of cross-layer digital twin collaboration. Consequently, the model compression for generative AI-enabled wireless network digital twins, as well as the wireless federated learning techniques, are demanding research issues for effective collaboration, especially within distributed wireless networks.}


\subsection{Privacy and Security of Network Digital Twin}
There are numerous 6G application scenarios that involve sensitive data, such as multisensory extended reality, sensing, and applications for business. In such cases, applications are susceptible to data theft by malicious entities during communication. Unfortunately, the establishment and physical-digital synchronization for digital twins demand a substantial volume of communication data. Therefore, it is essential to enhance the security and privacy of communications between the physical network and its digital twin. Moreover, reinforcing defense mechanisms against potential attacks and safeguarding the confidentiality of digital twins throughout \textcolor{black}{modeling} and synchronization processes are also imperative. The inherent stochastic nature of generative AI presents substantial potentials for addressing the privacy and security concerns, in terms of the encryption of valuable information, as well as the generation of numerous indistinguishable synthetic data points to protect privacy.

\subsection{Intelligent Deployment of Network Digital Twin}
The forthcoming 6G network is characterized in part by extensive 3D global coverage \cite{ITU2023}. Therefore, 6G wireless network digital twins will be deployed in either a distributed manner, at the end-user devices, or centrally within a cloud infrastructure. While distributed deployment offers the benefit of reduced synchronization latency, it suffers from limited computational capacity and increased management complexity, particularly for compute-intensive generative models. \textcolor{black}{On the other hand, a centralized deployment leverages high-performance computing and storage resources, at the cost of higher latency and weakened reliability. Consequently, a comprehensive and quantitative investigation is necessary to assess the pros and cons across various deployment strategies. It is a promising avenue for researchers to explore trade-offs between various deployment methods. In addition, the lack of standardized assessment criteria for network digital twins poses significant challenges in this domain, along with the difficulty in obtaining high-quality datasets for deploying network digital twins.}

\section{Conclusion}

\textcolor{black}{The digital twin technology holds substantial potential for the emulation, evaluation, and optimization of wireless networks. The forthcoming 6G era presents new challenges for wireless network digital twins. This study carries out a prospective analysis of key requirements for 6G wireless network digital twins and leverages cutting-edge generative AI technologies to meet the demands. We also propose a hierarchical generative AI-enabled network digital twin along with a typical use case to showcase the distinct benefits of generative AI-enabled digital twins. Potential future directions for generative AI-enabled wireless network digital twins are also discussed.}

\section*{Acknowledgments}

This work was supported in part by the National Key Research and Development Program under Grant 2020YFB1806608

\begin{IEEEbiographynophoto}{Zhenyu Tao}
 received his B.S. degree from the School of Information Science and Engineering, Southeast University in 2022, where he is currently pursuing a Ph.D. degree in information and communication engineering. His research interests include wireless network digital twins and machine learning in wireless networks.
\end{IEEEbiographynophoto}

\begin{IEEEbiographynophoto}{Wei Xu}[S'07, M'09, SM'15] received his Ph.D. degrees in communication and information engineering from Southeast University, Nanjing, China in 2009. Between 2009 and 2010, he was a Post-Doctoral Research Fellow at the University of Victoria, Canada. He is currently a Professor at Southeast University. He served as an Editor for \textsc{IEEE Transactions on Communications} from 2018 to 2023. He is currently an Area Editor of \textsc{IEEE Communications Letters}. He is a Fellow of IET. His research interests include information theory, signal processing, and machine learning for wireless communications.
\end{IEEEbiographynophoto} 

\begin{IEEEbiographynophoto}{Yongming Huang} [M’10, SM’17]  has been a faculty member with the National Mobile Communications Research Laboratory, Southeast University, Nanjing, China, where he is currently a Full Professor. He also serves as the Director of the Pervasive Communication Research Center, Purple Mountain Laboratories, Nanjing, China. His current research interests include intelligent 5G/6G mobile communications and mmWave wireless communications.
\end{IEEEbiographynophoto}

\begin{IEEEbiographynophoto}{Xiaoyun Wang} received the B.S. and M.S. degrees in electrical engineering from the Beijing University of Posts and Telecommunications, Beijing, China, in 1991 and 2002, respectively. She is Chief Scientist of China Mobile Communications Group Corporation currently. Her research interests include 6G network architecture, integrated architecture of computing and networking, network intelligence, etc.
\end{IEEEbiographynophoto}

\begin{IEEEbiographynophoto}{XiaoHu You} [F'11] has been working with National Mobile Communications Research Laboratory at Southeast University, where he holds the rank of director and chief professor. Now he is also the director and chief scientist of Purple Mountain Laboratory, and the deputy director of Pengcheng Laboratory. His research interests include broadband wireless transmission, mobile communication system, advanced signal processing and its applications. He was selected as IEEE Fellow in 2011. He won the Tan Kah Kee Science Award in 2014 and IET Achievement Medal in 2021. He was awarded as Academician of China Academy of Science in 2023.
\end{IEEEbiographynophoto}

\vfill

\end{document}